\documentclass[12pt]{article}
\usepackage{amsmath}
\usepackage{graphicx}
\usepackage{enumerate}
\usepackage{natbib}
\usepackage{url}

\usepackage{A_command}

\usepackage{bm}
\usepackage{url}
\usepackage{graphicx}
\usepackage{amsmath, amsthm}
\usepackage{setspace} 
\usepackage{natbib}
\usepackage{color}

\newcommand{\blind}{1}

\usepackage{capt-of}
\usepackage{booktabs}
\usepackage{varwidth}
\newsavebox\tmpbox

\usepackage{caption}
\usepackage{subcaption}

\usepackage{tikz}
\usetikzlibrary{shapes.geometric, arrows}
\usetikzlibrary{shapes.geometric, fit}
\usetikzlibrary{positioning}
\usetikzlibrary{arrows,calc}
\tikzset{
	>=stealth',
	punkt/.style={
		rectangle,
		rounded corners,
		draw=black, very thick,
		text width=5.5em, 
		minimum height=3em,
		text centered},
	pil/.style={
		->,
		thick,
		shorten <=2pt,
		shorten >=2pt,}
}

\definecolor{ydcolor}{RGB}{79, 51, 255}

\theoremstyle{definition}

\newtheorem{theorem}{Theorem}

\author{}
\date{}

\begin{document}

	\def\spacingset#1{\renewcommand{\baselinestretch}%
		{#1}\small\normalsize} \spacingset{1}
	
	
	\if1\blind
	{
		\title{\bf Bayesian Model Calibration and Sensitivity Analysis for Oscillating Biological Experiments}
		\author{Youngdeok Hwang \hspace{.2cm}\\
			Paul H. Chook Department of Information Systems and Statistics\\
			Baruch College, City University of New York\\
			Hang J. Kim\\
			Division of Statistics and Data Science, University of Cincinnati \\
			Won Chang \\
			Department of Statistics, Seoul National University\\
			Institute for Data Innovation in Science, Seoul National University\\
			Christian Hong \\
			Department of Pharmacology and Systems Physiology, 
			University of Cincinnati \\
			and \\
			Steven N. MacEachern\\
			Department of Statistics, The Ohio State University
		}
		\maketitle
	} \fi
	
	\if0\blind
	{
		\bigskip
		\bigskip
		\bigskip
		\begin{center}
			{\LARGE\bf Bayesian Model Calibration and Sensitivity Analysis for Oscillating Biological Experiments}
		\end{center}
		\medskip
	} \fi
	
	\bigskip
	\begin{abstract}
		Understanding the oscillating behaviors that govern organisms' internal biological processes requires interdisciplinary efforts combining both biological and computer experiments, as the latter can complement the former by simulating  perturbed conditions with higher resolution. 
		Harmonizing the two types of experiment, 
		however, poses significant statistical challenges due to 
		identifiability issues, numerical instability, and ill behavior in high dimension. This article devises a new Bayesian calibration framework for oscillating biochemical models. The proposed Bayesian model is estimated relying on an advanced Markov chain Monte Carlo (MCMC) technique which can efficiently infer the parameter values that match the simulated and observed oscillatory processes. Also proposed is an approach to sensitivity analysis based on the intervention posterior. This approach measures the influence of individual parameters on the target process  by using the obtained MCMC samples as a computational tool. 
		The proposed framework is illustrated with circadian oscillations observed in a  filamentous fungus, \textit{Neurospora crassa.}
	\end{abstract}
	
	\noindent%
	\vfill
	
	\newpage
	\spacingset{1.9} 
	
	
	\section{Introduction} \label{sec:introduction}
	Periodicity is a common characteristic of biological systems. Organisms demonstrate a diverse range of periodic behaviors that regulate distinct molecular, cellular, and physiological phenotypes such as the cell cycle, yeast metabolic cycle, segmentation clock in vertebrates, and circadian rhythms
	\citep{tu2005logic, Corsietal2018}. 
	Studying the molecular mechanisms behind these oscillators is often enhanced by combining both 
	physical experiments and computer models, 
	as the computer models allow us to complement the physical experiments by simulating numerous perturbed conditions at a higher resolution 
	\citep{gallego2006opposite, tsai2008robust}.
	
	Computer experiments are widely used to study complex physical processes in various applications in science and engineering \citep{santner2003design}.
	Many of these experiments are based on a computer model which consists of a set of differential equations, indexed by parameters and solved by numerical methods. Development of such a model is a major undertaking, with the goal of creating a family of models that is accurate (although imperfect) for some parameter values.  Determination of plausible sets of parameters that produce outcomes similar to those observed in the physical experiment is called \textit{calibration}.  A sound, calibrated model is said to be mature.   
	Having a mature model is particularly crucial for counterfactual scenario analyses where the conditions depart from the current steady-state and a system may become unstable.  For example, climate projections under various future CO$_2$ forcing scenarios require a rigorous parametric uncertainty quantification of the climate system \citep{chang2014fast}. 
	
	Since the seminal work of \cite{kennedy2001bayesian}, computer model calibration has been a major research topic.  \cite{higdon2008computer} introduce a basis representation  approach, and \cite{chang2014fast} reformulate this approach using dimension reduction to enable faster computing.  \cite{gramacy2015calibrating} focus on large data sets. \cite{chang2015binary} and \cite{sung2020calibration} consider binary responses, and \cite{hwang2018statistical} directly incorporate equations from physics within the calibration framework. There is an interest beyond the statistics community;  \cite{karagiannis2017bayesian}, for example, propose a Bayesian approach from the computational physics perspective. 
	
	Our work belongs to the broad Bayesian framework of \cite{kennedy2001bayesian}. Despite their overarching perspective and the successful applications that followed, direct application of their framework to our study faces multiple challenges. First, their formulation presumes that the output is expensive and scarce, yet well-behaved. When the computer model is affordable but shows highly erratic behavior in the input space, an alternative structure is needed for accurate calibration. Second,  \cite{kennedy2001bayesian} assume that the outcomes of
	physical experiments and computer models, 
	as well as their discrepancy, all lie in the same space. Such an assumption allows a direct evaluation of the likelihood in a Bayesian setting or the objective function in a frequentist setting \citep[e.g.,][]{tuo2015efficient}. In our motivating example, however, the outputs from both 
	physical experiments and the computer model 
	are cyclic and oscillating, which requires a customized measure for the discrepancy. A computer model run is deemed successful when its resulting curve has similar periodicity and amplitude with the curves from the physical experiments, allowing lateral (phase) shifts. 
	
	There are three major contributions in our work. First, our modeling approach characterizes the periodicity and amplitude of the  outcome through a harmonic basis representation, so that it captures the oscillating nature of the experiments. 
	Second, we introduce a Markov chain Monte Carlo (MCMC) approach built on the generalized multiset sampler \citep[GMSS,][]{kim2015generalized} to efficiently evaluate the posterior distribution. 
	Supplemented by a computational approach to find the instrumental densities for the GMSS, 
	the method provides an efficient solution to overcome the challenges in exploring a high-dimensional and multi-modal posterior density. It incorporates the instrumental density within the computing scheme in a unique fashion. Third, we propose an intuitive method for sensitivity analysis to handle parameter uncertainty. 
	Our framework is illustrated with a case study on the circadian cycle of a living organism that consists of physical experiment measurements and a corresponding computer model. Together, these three components contribute to solving a complex scientific problem, presenting a thorough case study.
	
	The remainder of the paper is organized as follows. Section \ref{sec:background} describes the basics of oscillating biochemical experiments and the current practice for analysis. Section \ref{sec:model} describes the proposed model and methodology. Section \ref{sec:intervention-posterior} proposes the intervention posterior approach for sensitivity analysis. Section \ref{sec:case-study} presents the case study for an analysis of the  three-variable biochemical oscillator using the proposed methodology. We conclude with some remarks and discussion in Section \ref{sec:discussion}.
	
	\vspace{-2em}
	\section{Scientific Problem}
	\label{sec:background}
	
	\subsection{Oscillating Biochemical Experiment: Circadian Cycle} 
	\label{subsec:scientific-problem}
	Circadian rhythms are events that recur with a period of about 24 hours. 
	The circadian clock governs periodic behaviors that respond to external cues such as the light-dark cycle  
	and aligns the internal clock to the external environment to optimize an organisms' function. Misalignment of the internal  clock and the external environment increases the risk of sleep disorders, as well as cardiovascular and metabolic diseases. 
	Thus, it is critical to understand the fundamental mechanisms of circadian rhythms and their signaling pathways to other processes \citep{bell2005circadian}.
	
\begin{figure}
		\centering
		\includegraphics[width=0.45\textwidth]{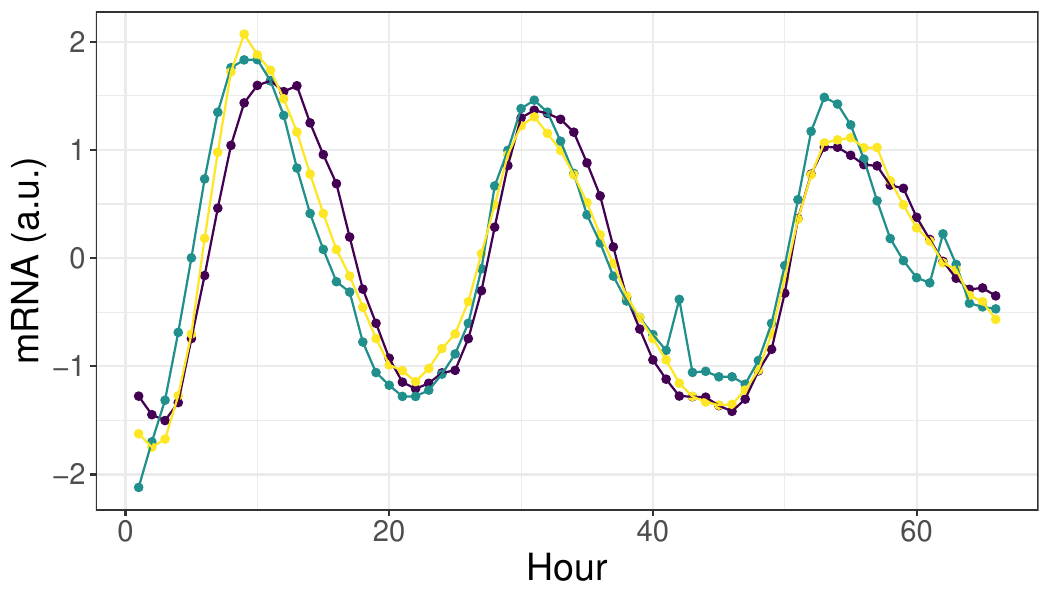}
		\caption{Bioluminescence measurements from a physical experiment with three replicates.}
		\label{fig:physical-experiment}
	\end{figure}

	In this work, we use the circadian cycle of a filamentous fungus \emph{Neurospora crassa}.  \emph{N.crassa} has been a model organism for understanding circadian clocks since the 1950s, when its physiological output became tractable to biologists \citep{dunlap2017making}. Physical experiments in this study produce indirect measurements of the gene expression activity of the gene named \emph{frequency (\textit{frq})}, which is known to regulate the circadian cycle of \emph{N.crassa}. The activity of \textit{frq}  expression is measured using bioluminescence assays that detect the activity of luciferase driven by the \textit{frq} promoter \citep{gooch2014kinetic}. Bioluminescence is detected by a software-controlled camera collecting the measurements for 10 minutes every hour. Figure \ref{fig:physical-experiment} depicts the three replicates from physical experiments we use in this study, where we observe endogenous cycles of approximately 21--22 hours.
	
\begin{figure}
		\centering
		\tikzstyle{line} = [draw, -latex']
		\begin{tikzpicture}[scale=0.6, transform shape]
			\node[punkt](M){$y$};
			\node[punkt, right = 2cm of M](P){$w$};
			\node[punkt, right = 2cm of P](Pp){$z$};
			\node[draw=none,fill=none, right = 2cm of Pp](Rhollow){};
			\node[draw=none,fill=none, left = 2cm of M](Lhollow){};
			\node[draw=none,fill=none, below = 1cm of M](B1){};
			\node[draw=none,fill=none, below = 1cm of P](B2){};
			\path (Pp.north) edge[out=135, in=45, ->]  node[anchor=south, above]{  $\theta_8$}    node[anchor=east, above]{\small } (M.north); 
			\draw[-latex] (Lhollow.east) -- node[auto,] { c} (M.west);
			\draw[-latex] (Lhollow.east) -- node[auto,below] {  } (M.west);
			\draw[-latex] (M.east) -- node[auto,] { $\theta_2$} (P.west);
			\draw[-latex] ($(P.east)-(0,0.1)$) -- node[auto, below] {  $\theta_4$} ($(Pp.west) -(0,0.1)$);
			\draw[-latex] ($(Pp.west)+(0,0.1)$) -- node[auto,above] {  $\theta_6$} ($(P.east)+(0,0.1)$);
			\draw[-latex] (Pp.east) -- node[auto,] {  $\theta_5$} (Rhollow.west);
			\draw[-latex] (M.south) -- node[auto,] {  $\theta_1$} (B1.north);
			\draw[-latex] (P.south) -- node[auto,] {  $\theta_3$} (B2.north);
			\path (Pp) edge[out=-105, in=-75, ->]  node[anchor=south, above]{\small } node[anchor=east, below]{  $\theta_7, \theta_9$} ($0.5*(Pp.west)+0.5*(P.east) - (0,0.6)$); 
		\end{tikzpicture}
		\caption{\label{fig:overall-description} The schematic illustration of the oscillator experiment. }
		\centering
	\end{figure}
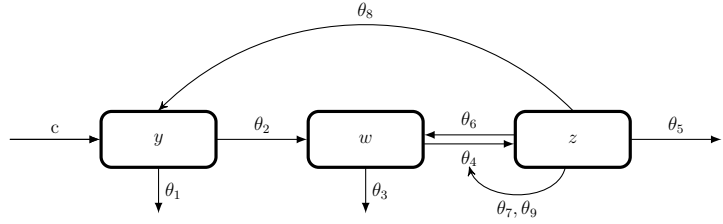

	As a counterpart of the physical experiments, we consider the mathematical model for the molecular mechanism of oscillators regulated by nonlinear dynamics, defined by a set of ordinary differential equations (ODE)  from \cite{caicedo2015robustness}.
	This simple model contains both negative and positive feedback loops that generate autonomous oscillations.  It consists of three state variables: \emph{frq} mRNA ($y$), protein ($w$), and phosphorylated protein ($z$) as the main variables, where the system wiring is characterized by nine unknown parameters $\bmtheta = (\theta_1, \ldots, \theta_9)$ that are related to biological processes, such as the synthesis rate or the  threshold of critical concentration.  The parameter related to the transcription rate of $y$ is fixed at $c$. 
	\cite{caicedo2015robustness} wrote this dynamical system as
	\begin{eqnarray}
		\frac{dy}{dt} & = & \frac{c}{1 + (z/\theta_8)^8} - \theta_1 y, \label{eqn:y-ode}\\ 
		\frac{dw}{dt} & = & \theta_2 y - (\theta_3 + \theta_4) w + \theta_6 z - \frac{\theta_7 w z^4}{\theta_9^4 + z^4}, \label{eqn:w-ode}\\
		\frac{dz}{dt} & = & \theta_4 w - (\theta_5 + \theta_6) z + \frac{\theta_7 w z^4}{\theta_9^2 + z^4}. \label{eqn:z-ode} 
	\end{eqnarray}
	In this system, the \emph{frq} mRNA ($y$) is translated into protein ($w$), and this protein is transformed
	into the end product ($z$). 
	Figure \ref{fig:overall-description} presents a schematic illustration of this system, where arrows represent how each sub-process is involved in the process.
	For example, $\theta_1$ in \eqref{eqn:y-ode} is the degradation rate for $y$, and $\theta_2$ in \eqref{eqn:w-ode} is the translation rate from $y$ to $w$. From Figure \ref{fig:overall-description} and equations \eqref{eqn:y-ode}-\eqref{eqn:z-ode}, it can be seen that  the system has self-regulating feedback loops;  
	for example, a high level of $y$ leads to a higher level of $w$ through $\theta_2$, then a higher level of $z$ through $\theta_4$, but the increase in $z$ (modulated by $\theta_8$) hinders the transcription of $y$. The amount of mRNA in the computer experiment, $y$, is to be matched with the bioluminescence measurements from physical experiments for model calibration. 
	
	In our specific scientific objective, the (true but unknown) circadian model assumes ``stationary'' periodic oscillations for the {\it frq} mRNA, which has triplicate experimental measurements in our analysis. To reach a stationary periodic solution from the ODEs in (1)--(3), the ODE simulator is run for 500 time steps after setting initial values of the states, i.e., $(y^{(0)},w^{(0)},z^{(0)})$ in \eqref{eqn:y-ode}--\eqref{eqn:z-ode}, 
	and we use the last 66-hour portion after discarding the first 434 time steps as equilibrium steps \citep{ormeno2024convergence}. Because biologists are interested in the periodic circadian rhythm, the model outcome in the early stage needs to be discarded and not used for the analysis. These equilibrium steps cause substantial variability in the phase of the ODE output.
	

	\subsection{Current Practices in Mathematical Biology}
	\label{subsec:CurrentPractice}
	
	When studying a phenomenon using both computer and physical experiments, it is imperative to harmonize the two sources by tuning the parameters of the computer model to match physical measurements.  Typical analyses begin with an initial parameter value from the existing literature or an experimenter's hunch, followed by a search for a \emph{better} parameter value, perhaps by trial-and-error or with a more formal yet heuristic search algorithm, until a \emph{reasonable} value $\bmtheta_0$ is found. This search may or may not be directly tied to the physical experiment; it often relies on visual examination to find a subjective match of qualitative features of the computer model output and physical experiment measurements. 
	Optimization algorithms, such as simulated annealing \citep{kirkpatrick1984optimization} 
	and multiple shooting \citep{peifer2007parameter}, figure prominently in these searches. 
	A post-hoc sensitivity analysis is often conducted 
	once $\bmtheta_0$ is found. The sensitivity analysis  seeks to provide insight into qualitative or quantitative changes in the system behavior, as the parameters deviate from $\bmtheta_0$ \citep{caicedo2015robustness, liu2019frq},  yet it is often limited to a slight perturbation \citep{landa2022phage}.
	
	These calibration practices often suffer from slow convergence, local maxima, and sensitivity to tuning parameters. They also lack a suitable measure of the inherent uncertainty in estimating the model parameters and sensitivity analysis thereafter, yet these limitations are  not clearly addressed in the mathematical biology literature \citep[e.g.,][]{bellman2018modeling}. 
	
	\vspace{-1.7em}
	\section{Model Formulation and Posterior Inference} \label{sec:model} 
	
	Many parameters in computer experiments are unknown to the experimenters, and the computer experiments rarely fully reflect reality, causing some degree of bias. Since the seminal work of \cite{kennedy2001bayesian}, Bayesian approaches have been widely used for computer model calibration, and inference problems have been cast in the Kennedy-O'Hagan framework. In this section, we present our modeling approach to integrate the physical and computer experiments in a Bayesian fashion. 
	
	\vspace{-1.5em}
	\subsection{Bayesian Hierarchical Model with Harmonic Basis Representation}
	
	Let $\bm{y}_i^{F} = (y_{i0}^{F}, \ldots, y_{i,T-1}^{F})^{\top}$ denote the field data measured from the $i$-th physical experiment. To incorporate the strong periodicity of the observed data, we formulate
	\begin{align}  \label{eqn:basis-form}
		& y_{it}^{F} = \sum_{k=1}^{K}  a_{ik}  \cos \left(\frac{2 \pi k  t}{T} \right)  +  \sum_{k=1}^{K}  b_{ik} \sin\left(  \frac{2 \pi k  t}{T} \right)  + \varepsilon_{it}, 
	\end{align}
	for $ i=1,\ldots,n$ and  $t=0,\ldots,T-1$,  where $\varepsilon_{it}$ is assumed to be an independent error from $N(0, \sigma_i^2)$.  
	This formulation uses a truncated Fourier series
	to characterize the physical experiment, we consider $s_{ik} = a_{ik}^2 + b_{ik}^2$, the power spectrum of the $k$-th frequency component, which represents the magnitude of periodic behavior at the frequency of $k/T$ in the $i$-th observed  time series \citep{hartley1949tests}. 
	For example, the data sets presented in Figure \ref{fig:physical-experiment} have $T=66$, so $s_{i1}$ represents the magnitude of the periodic behavior that completes its cycle once every $66$ hours, $s_{i2}$ every 33 hours, and so on.
	
	For the computer model, $\{ y_{0} (\bmtheta), \ldots, y_{T-1} (\bmtheta) \}$ denote outcomes executed with the input parameters $\bmtheta = (\theta_1,\ldots,\theta_p)$. Then, we approximate the computer model outcomes by 
	\vspace{-1em}
	\begin{align}
		y_{t}(\bmtheta) =  \sum_{k=1}^{K}  \alpha_k(\bmtheta)  \cos\left(  \frac{2 \pi k  t}{T} \right)  +  \sum_{k=1}^{K}  \beta_k(\bmtheta) \sin\left(  \frac{2 \pi k  t}{T} \right) + \varepsilon_{t}^M,
		\label{eqn:mm-basis-form}
	\end{align} 
	\noindent
	for $t=0,\ldots,T-1$ where $\varepsilon_{t}^M$ is an approximation error due to the harmonic basis representation.
	We let $\tilde{\alpha}_k(\bmtheta)$ and $\tilde{\beta}_k(\bmtheta)$ denote the minimizer of the sum of squared distances between $ y_{t}(\bmtheta)$ and the fitted curve and define  
	$\lambda_k(\bmtheta) = \tilde{\alpha}_k^2(\bmtheta) + \tilde{\beta}_k^2(\bmtheta)$ similarly to $s_{ik}$ from the field data. 
	Note that $\lambda_k(\bmtheta)$'s are deterministic functions of $\bmtheta$ with these parameterizations.

	The formulations in \eqref{eqn:basis-form} and \eqref{eqn:mm-basis-form} are harmonic basis representations. A suite of different basis functions has been employed in the computer experiments literature, including wavelets \citep{bayarri2007computer}, principal components \citep{higdon2008computer,chang2014fast,chang2015binary}, and Gaussian kernels \citep{bhat2013inferring}. 
	To the best of our knowledge, however, harmonic basis functions have not been utilized to characterize the periodicity of the model outcome in computer model calibration literature.
	
	Following the Kennedy-O'Hagan framework, we connect the physical data and computer model outputs using the underlying spectrum by $s_{ik} = \lambda_k(\bmtheta) + \delta_{ik}$ with the  model discrepancy $\delta_{ik} \sim N(0, \tau^2)$ for $i=1,\ldots,n$ and $k=1,\ldots,K$. To complete the  hierarchical structure for calibration, adding prior distributions for $\bmtheta$, { $\sigma_i^2$}, and $\tau^2$ gives the following hierarchical Bayesian model 
	
	\vspace{-3em}
	\begin{align}
		& [y_{it}^F | \bma_i, \bmb_i, \sigma_i^2]  \sim N \left( \sum_{k=1}^{K}  a_{ik}  \cos \left(\frac{2 \pi k  t}{T} \right)  +  \sum_{k=1}^{K}  b_{ik} \sin\left(  \frac{2 \pi k  t}{T} \right), \sigma_i^2 \right),\nonumber\\
		& [s_{ik}|\lambda_k(\bmtheta), \tau^2] \sim N^+(\lambda_k(\bmtheta), \tau^2), ~ [\theta_j] \sim \text{IG} (\nu_\theta, \psi_\theta), ~
		[\sigma_i^2] \sim \text{IG} (\nu_\sigma, \psi_\sigma), ~
		[\tau^2] \sim \text{IG} (\nu_\tau, \psi_\tau),   \label{eqn:BayesianModel} 
	\end{align} 
	
	\vspace{-2em} \noindent
	for $i=1,\ldots,n$, $j=1,\ldots,p$, $k=1,\ldots,K$, and $t=0,\ldots,T-1$ 
	where $\bma_i=(a_{i1},\ldots,a_{iK})$, $\bmb_i=(b_{i1},\ldots,b_{iK})$, and $N^+$ and $\text{IG}$ denote the truncated normal distribution above zero and the inverse gamma distribution, respectively. Note that we do not explicitly specify the priors for $\lambda_k(\bm{\theta})$ which is defined based on those of $\bm{\theta}$. The posterior distribution is evaluated with the model spectrum $\lambda_k(\bmtheta)$, 
	so the computer model needs to be executed for each MCMC iterate. 
	Our approach is best suited for computer experiments that are computationally affordable \citep[e.g.,][]{lee2020fast} as described in Section \ref{sec:background}. 
	
	Pairing $\lambda_k(\bmtheta)$'s and $s_{ik}$ within the Kennedy-O'Hagan framework, rather than separately pairing $(a_{ik}, \tilde{\alpha}_k(\bmtheta))$ and $(b_{ik}, \tilde{\beta}_k(\bmtheta))$, offers a robust formulation for calculation of the likelihood in the simulation model. The simulation model produces an artificial horizontal shift of undetermined length at the beginning of the outcome. As such, either $\tilde{\alpha}_k(\bmtheta)$ or $\tilde{\beta}_k(\bmtheta)$ bears the dominant periodic behavior depending on the horizontal shift. Thus, individually pairing the two basis coefficients at each $k$ may can result in misleading likelihood evaluations. In contrast, in our calibration, two $\bmtheta$'s that produce outcomes that only differ in phase (e.g., an exact two hour shift to the right but no change in amplitude) may correspond to very different sets of $\tilde{\alpha}_k(\bmtheta)$ and $\tilde{\beta}_k(\bmtheta)$, yet yield identical likelihoods. Modeling $\lambda_k$ addresses this issue, by taking into account both frequency and amplitude.  Furthermore, this is a cost-effective approach, particularly given that the simulator needs to be run extensively over MCMC. For more general discussion on the issues of model calibration with functional data, see \cite{francom2023elastic}.
	
	Our approach focuses on matching the frequency and amplitude of computer model outputs and observed field data, without an additional discrepancy term in the time domain. Since there is no clear additional variation found other than the cyclic behaviors in the observed data, omitting a systematic discrepancy term in the time domain does not cause an issue in our application. If an additional discrepancy term is needed for better prediction or to incorporate prior knowledge on data-model discrepancies, one might consider other approaches including those of \cite{liu2009modularization}, \cite{francom2019inferring}, \cite{gu2019jointly}, and \cite{salter2019uncertainty}. In  particular, if a Gaussian process is used for the discrepancy, the approach in \cite{gu2019jointly} can provide a robust prior choice that avoids possible inferential issues, including nonidentifiability, between the input parameters and the data-model discrepancy.

	The modeling structure in \eqref{eqn:basis-form}-\eqref{eqn:BayesianModel} encompasses the periodic nature of the computer model in \eqref{eqn:y-ode}-\eqref{eqn:z-ode}.  However, the complex interactions between the model parameters often make one stimulus dominate the entire feedback loop process, rendering the regulating cycle inoperative. 
	Accordingly, the model produces a periodic solution only in a small portion of the parameter space which consists of many narrow sub-regions that are disconnected from each other and are scattered over the high-dimensional parameter space.  As a consequence, the posterior density is very low in the majority of the input space, posing a critical challenge for a standard MCMC algorithm. To address this challenge, we propose a computational solution with two main building blocks as described in the following sections.

	\vspace{-1em}
	\subsection{Posterior Estimation by the Generalized Multiset Sampler} \label{subsec:gmss}
	The presence of multiple local maxima and narrow high-density regions causes a standard MCMC sampler to get trapped in a local maximum or to show prohibitively slow convergence and mixing.
	Such issues have long been recognized as a major challenge in applications of MCMC sampling. A number of advanced Monte Carlo algorithms have been proposed to alleviate this issue, such as the tempering algorithm and its variants \citep{swendsen1986replica, geyer1991markov, marinari1992simulated}, the multiple-try Metropolis algorithm \citep{liu2000multiple}, and the multiset sampler \citep{leman2009multiset,kim2015generalized}.
	
	The joint posterior distribution of all unknown parameters in our hierarchical Bayesian model in \eqref{eqn:BayesianModel} is 
	$f( \bma, \bmb, \bmtheta, \bmsigma^2, \tau^2 | \bmy^F )
	\propto 
	\left[ \prod_{i=1}^n \left\{ \prod_{t=0}^{T-1} f( y_{it}^F | \bma_i, \bmb_i, \sigma_i^2 ) \prod_{k=1}^K f( s_{ik} | \lambda_k(\bmtheta), \tau^2 ) \right\} \right]$
	$ f(\bmtheta) f(\bmsigma^2) f(\tau^2) $
	where $\bma = \{ \bma_1, \ldots, \bma_n \}$, $\bmb = \{ \bmb_1, \ldots, \bmb_n \}$, $\bmsigma^2= \{ \sigma_1^2, \ldots, \sigma_n^2 \}$, and $\bmy^F = \{ \bmy_1^F, \ldots, \bmy_n^F \}$.
	A posterior inference approach using standard Metropolis-within-Gibbs is given in the Supplement with detailed steps.  This algorithm fails to explore the posterior distribution of our model parameters.
	
	We address these challenges by adopting an advanced MCMC method called the generalized multiset sampler. First, we introduce the \emph{multiset sampler} originally proposed by \cite{leman2009multiset} and then describe the \emph{generalized} multiset sampler \citep{kim2015generalized}.  
	The multiset sampler is an MCMC algorithm originally designed to make inference for a multimodal distribution. 
	Suppose that $\bmx$ is a set of observations and $\bmtheta \in {\Omega}$
	is a vector of parameters whose posterior density is believed to be multimodal. 
	The central idea of the multiset sampler is 
	to define a multiset $\bmTheta = \{ \bmtheta_1, \ldots, \bmtheta_M \}$ for the model parameter $\bmtheta$, 
	draw  MCMC samples of the multiset, and make  an inference about $\bmtheta$ by statistically fusing the multiple copies of $\bmtheta_m$ in the multiset, instead of directly sampling the original parameter $\bmtheta$.
	When the target posterior density is $f(\bmtheta | \bmx)$, 
	the multiset sampler updates the parameters with its multiset \emph{sampling} distribution $\pi(\bmTheta|\bmx)$, defined as $\pi(\bmTheta|\bmx) \propto \sum_{m=1}^M f(\bmtheta_m|\bmx)$.  
	At iteration $b=1,\ldots,B$, 
	the MCMC updates the $m$-th multiset element $\bmtheta_m$ given the current values of $(\bmtheta_1,\ldots,\bmtheta_M)$ via the Metropolis-Hastings algorithm: propose $\bmtheta_m^q \sim q(\bmtheta_m^q|\bmtheta_m)$ and accept as $\bmtheta_m^{(b)}=\bmtheta_m^q$ with the acceptance probability,  $\min\{1,\alpha_m^{(b)}\}$, where
	
	\vspace{-1.0em}
	\[ \alpha_m^{(b)} = 
	\frac{f(\bmtheta_m^q|\bmx) + \sum_{l \neq m} f(\bmtheta_l|\bmx) }{f(\bmtheta_m|\bmx) + \sum_{l \neq m} f(\bmtheta_l|\bmx)}
	\frac{q(\bmtheta_m|\bmtheta_m^q)}{q(\bmtheta_m^q|\bmtheta_m)}. \]
	
	\noindent 
	This accept-reject update
	shows how the multiset sampler avoids getting stuck in a local mode. 
	For simplicity of illustration, let us assume that the proposal distribution is symmetric, so that $q(\bmtheta_m|\bmtheta_m^q)/q(\bmtheta_m^q|\bmtheta_m)$ is always one, 
	although this is not necessary in practice. 
	If $f(\bmtheta_m^q|\bmx)$ and $f(\bmtheta_m|\bmx)$ are relatively large compared to $f(\bmtheta_l|\bmx)$ for all $l \neq m$, the value of $\alpha$ is close to 
	$f(\bmtheta_m^q|\bmx) / f(\bmtheta_m|\bmx)$, 
	so the Markov chain behaves similarly to a standard Metropolis-Hastings random walk, exploring the original target distribution $f(\bmtheta|\bmx)$. We call this multiset element the \emph{leading} element. 
	Meanwhile, if $\bmtheta_m$ is associated with a relatively small value of $f(\bmtheta_m|\bmx)$, 
	it contributes little to 
	$f(\bmtheta_m|\bmx) + \sum_{l \neq m} f(\bmtheta_l|\bmx)$ 
	and hence the acceptance probability for the proposed $\bmtheta_m^q$ becomes nearly one,  
	regardless of the suggested $\bmtheta_m^q$ value. 
	This \emph{non-leading} element of the multiset sampler is let loose in $\Omega$  and travels freely, without getting stuck in a local mode.
	
	The generalized multiset sampler \citep[GMSS,][]{kim2015generalized} 
	refines the idea of the multiset sampler by explicitly linking the sampling distribution to the target distribution and by using an auxiliary density to guide the movement of the Markov chain toward plausible areas. Drawing on results from importance sampling, it defines the multiset sampling distribution by 
	\vspace{-1.5em}
	\begin{equation}
		\pi(\bmTheta|\bmx) = \frac{1}{M} \sum_{m=1}^M f(\bmtheta_m|\bmx) \prod_{l \neq m} g_l(\bmtheta_l),
		\label{eq:GMSS}
	\end{equation}
	\vspace{-2.5em}
	
	\noindent
	where $g_l$ is a density with the same support as $\bmtheta$, which is referred to as the {\it instrumental density}. 
	In this general framework, the original multiset sampler of  \cite{leman2009multiset} is a special case of \eqref{eq:GMSS} where $g_l(\bmtheta_l)$ is a uniform distribution on a bounded support.
	With some algebra, the marginal sampling density of each multiset element is derived 
	to be $\pi(\bmtheta_m|\bmx) = (1/M)f(\bmtheta_m|\bmx) + \left\{1 - (1/M) \right\} g_m(\bmtheta_m)$
	for $m=1,\ldots,M$,
	which is a weighted sum of the target density $f(\bmtheta|\bmx)$ and the instrumental density $g_m(\bmtheta)$. 
	The instrumental density is a tuning density that gently herds the sampler toward a promising area. 
	If $g_m(\bmtheta)$ were chosen to be precisely $f(\bmtheta_m|\bmx)$, 
	the marginal sampling distribution $\pi(\bmtheta_m|\bmx)$ becomes the target distribution $f(\bmtheta|\bmx)$ which guarantees the most efficient sampling scheme.  In practice,  
	even crude information on the target density can be useful to form an instrumental density. 
	For any integrable function of the parameters, $h(\bmtheta)$, inference on $h(\bmtheta)$ under  the target posterior $f(\bmtheta|\bmx)$ is made from the GMSS sampling distribution $\pi(\bmTheta|\bmx)$ by Theorem 1 in \cite{kim2015generalized}, which is restated below:
	
	\begin{theorem}
		Define a set of weights 
		\[
		w_m = \frac{f(\bmtheta_m|\bmx) \prod_{l \neq m} g_l(\bmtheta_l)}{\sum_{m'=1}^M f(\bmtheta_{m'}|\bmx) \prod_{l' \neq m'} g_{l'}(\bmtheta_{l'})} \ \text{ for } m=1,\ldots,M.
		\]
		\noindent 
		Then, for any integrable function $h$, $\E_f \left\{ \left. h( \bmtheta ) \right| \bmx \right\} = \E_\pi \left\{ \left. \sum_{m=1}^M w_m h( \bmtheta_m ) \right| \bmx \right\}$.
	\end{theorem}
	\noindent
	
	This theorem implies that the importance of elements in each multiset is reflected in the final inference via weight $w_m$.
	The conditional density under the GMSS sampling distribution $\pi$ is found by the following Metropolis-within-Gibbs steps. For each MCMC iteration $b=1,\ldots,B$, and
	for component $m=1,\ldots,M$, propose $\bmtheta_m^q \sim q(\bmtheta_m^q|\bmtheta_m)$ and let $\bmtheta_m^{(b)} = \bmtheta_m^q$ with probability $\min(1,\alpha_m^{(b)})$, where
	$\alpha_m^{(b)} = 
	\{ \pi(\bmTheta_m^q|\bmx) / 
	\pi(\bmTheta_m|\bmx) \} 
	\{ q(\bmtheta_m|\bmtheta_m^q) /
	q(\bmtheta_m^q|\bmtheta_m) \} 
	$
	with the proposed multiset $\bmTheta_m^q = \{ \bmtheta_1, \ldots, \bmtheta_{m-1}, \bmtheta_m^q, \bmtheta_{m+1}, \ldots, \bmtheta_M \}$ and the current state $\bmTheta_m = \{ \bmtheta_1, \ldots, \bmtheta_M \}$. 
	Then, the formal estimator of any target posterior quantity for a function $h(\bmtheta )$ is computed with the GMSS samples by 
	
	\vspace{-1.5em}
	\begin{equation}
		\widehat{\E}_f \{ h(\bmtheta ) | \bmx \}
		= \frac{1}{B} \sum_{b=1}^{B} \sum_{m=1}^M w_m^{(b)} h(\bmtheta_m^{(b)}),
		\label{eq:GMSSestimation}
	\end{equation}
	where the set of weights at the $b$-th iteration is
	\noindent
	$
	w_m^{(b)} = 
	\{ f(\bmtheta_m^{(b)} |\bmx) \prod_{l \neq m} g_l(\bmtheta_l^{(b)}) \} / 
	\{ M \ \pi(\bmTheta^{(b)} |\bmx) \}
	$
	and $\bmTheta^{(b)} = \{ \bmtheta_1^{(b)}, \ldots, \bmtheta_M^{(b)} \}$.
	
	Using the instrumental densities, the GMSS makes an explicit link between the target distribution $f(\bmtheta |\bmx)$ and the sampling distribution $\pi(\bmTheta |\bmx)$  used in the MCMC. 
	The GMSS provides an unbiased estimator of the posterior mean $E_f \{ \bmtheta  | \bmy \}$ as well as the marginal (posterior) cumulative distribution function of $\theta$, an element in $\bmtheta$, or a posterior quantile by setting $h(\theta) = I[ \theta \le a]$ where $a$ denotes a fixed evaluation point in the support of $\theta$. We refer interested readers to \cite{kim2015generalized} for various estimands introduced in the examples. 
	The generalized formulation allows  the algorithm to be run on an unbounded parameter space $\Omega$, with
	room for instrumental densities to further improve efficiency. 
	With the computer model being affordable,  a unique opportunity is available
	to find a  set of instrumental densities by exploiting computational resources.  
	
	\vspace{-1.5em}
	\subsection{Prognostic Experiments to Find Instrumental Densities} 
	\label{subsec:prognostic-experiments}
	
	To find an instrumental density, $g_m(\bmtheta)$, where the model runs successfully only on a very narrow area, our approach first evaluates the likelihood function with numerous input parameter settings using  high-throughput computing \emph{before} initiating the MCMC computation. 
	With those evaluations, the input space is categorized into two groups: primary search areas with a better prospect of success, and secondary areas with a lesser prospect of success, respectively.  We call this indexing process {\em prognostic} experimentation.
	
	We assume that the parameter space is $(0,1]$ for each dimension of the parameter $\bmtheta$ after appropriate scaling and transformation. We first create a large design $\mathcal{D}$ of $N$ runs, $\bmtheta_1, \ldots, \bmtheta_N$, where each run is independently generated from $U(0,1]^p$, and execute the computer model with them to obtain $\bmy_1, \ldots, \bmy_N$, where $\bmy_i$ is the output associated with $\bmtheta_i$. Since each run is independently generated, they can be readily distributed over multiple machines. To define the instrumental density, we adopt an approach inspired by the orthogonal array-based space-filling design \citep[OASD,][]{tang1993orthogonal}. The OASD was originally developed to achieve variance reduction in mean estimates when a small number of computer experiments can be afforded. Our focus is, however, on systematically indexing the input space using the orthogonal array structure while making full use of computational resources. 
	
	Let $ \lceil \cdot \rceil$ denote the ceiling function.  
	We partition $(0,1]$ into $q$ segments of equal length, $(0,1/q], (1/q, 2/q], \ldots, ((q-1)/q, 1]$, and define $u_q(\theta) = \lceil \theta q \rceil, \theta \in (0,1]$, which maps the parameters from $(0,1]$ to $\{1,\ldots,q\}$,
	and $u_q(\bmtheta) = \left(u_q(\theta_1), \ldots, u_q(\theta_p) \right)$ with the element-wise mapping. With this mapping, any point in $(0,1]^p$ is associated with a level combination of the $q^p$ factorial design. 
	Consider $\mathcal{S} = \{1,\ldots, p\}$ and all possible subsets $\mathcal{S}_b \subset \mathcal{S}$ with $|\mathcal{S}_b| = d_0 \leq p$ for $b = 1,\ldots, \genfrac(){0pt}{2}{p}{d_0}$, where $d_0$ is the chosen subspace search size specified by the user. Let $\mathcal{H}_b = \{h_{b}^{(1)}, \ldots, h_{b}^{(q^{d_0})}\}$ denote the available level combinations for $\mathcal{S}_b$, and $\bmomega_{\mathcal{S}_b} \in (0,1]^{d_0}$ the values of $\bmomega$ in $\mathcal{S}_b$ for an $\bmomega \in (0,1]^{p}$. 
	
	Now choose $l_{\min}$ and $n_{\min}$, where the former is the minimum threshold for the evaluated likelihood in \eqref{eqn:BayesianModel}, denoted by $l(\cdot )$, and the latter the minimum number of successful runs for a design space to be designated as a primary search area. With chosen $l_{\min}$, 
	find $G_b(h) = \{ \bmomega \in \mathcal{D}: u_q( \bmomega_{\mathcal{S}_b} ) = h, l(\bmomega) > l_{\min}\}$
	for $b = 1,\ldots, \genfrac(){0pt}{2}{p}{d_0}$ and $h \in \mathcal{H}_b$. 
	If $|G_b(h)| > n_{\min}$, the search categorizes $\{ \bmomega \subset (0,1]^p: u_q( \bmomega ) = h\}$ as a high prospect area. That is, a level in $q^p$ factorial design is deemed a high prospect if it has more than $n_{\min}$ successful runs, and an area 
	that shares has $d_0$ or more columns with any high prospect area after $u_q(\bmtheta)$ mapping is designated as a primary search area. 
	Each element in $\mathcal{H}_b$ represents $q^{-d_0}$ fraction of $(0,1]^p$ space, yet $\mathcal{S}_{b}$ and $\mathcal{S}_{b'}$ with $b\neq b'$ consider a projection into different subspaces. 
	When $\mathcal{D}$ is a factorial design, every $h \in \mathcal{H}_b$ in projection onto each $b$ represents a $q^{p-d_0}$ factorial design of $\mathcal{S} \setminus \mathcal{S}_b$ \citep{hwang2016sliced}. The parameters in the algorithm, $q$ and $d_0$, can be chosen while analyzing the results from prognostic experiments based on empirical principles, such as hierarchy, sparsity, and heredity \citep{wu2011experiments}. 
	
	After the search process, we set the instrumental density as  
	$g_m(\bmtheta) = \rho_1$ for the high prospect area and $\rho_0$ for the low prospect area, for all $m$. The values of $\rho_0$ and $\rho_1$ are determined to reflect the relative importance of the two areas in guiding the multiset sample and  to be scaled appropriately so that $\int_{\Omega} g_m(\bmtheta) \diff \bmtheta = 1$.
	The obtained instrumental densities are a primitive density estimate, and set loose so that they gently guide non-leading  elements toward exploring 
	the high \emph{prospect} area, while the leading  element stays around a currently known high \emph{density} region.
	
	\vspace{-1.6em}
	\section{Sensitivity Analysis via Intervention Posterior}
	\label{sec:intervention-posterior}
	An important aspect of scientific discovery is to find what changes to the system are likely to result in substantive changes to its features of interest. 
	This question follows a chain of causality -- experimental manipulation, or \emph{intervention}, 
	changes the ODE parameters $\bmtheta$ to impact a selected feature of the biological process, given by $h(\bmtheta)$, a function of the  parameters of research interest. We wish to assess the impact of intervention without performing additional physical experiments.

	For a system with known parameters, the sensitivity of $h(\bmtheta)$ to a change in a single parameter, say $\theta_j$, would be captured by defining

	\vspace{-1.5em}
	\begin{equation} 
		\bmnu_{j,\alpha} = \bmnu_{j,\alpha}(\bmtheta) 
		= \left( \theta_{1}, \ldots, \theta_{j-1}, \alpha \theta_{j},  \theta_{j+1}, \ldots,
		\theta_{p} \right), \label{eqn:intervened-sample}
	\end{equation}
	\vspace{-2.5em}
	
	\noindent and then computing $h(\bmnu_{j,\alpha} )$ for a range of values of a scale parameter, $\alpha$.  All perturbations under consideration must result in perturbed parameter values that lie within the parameter space.  
	Alternatively, 
	one could consider the local sensitivity of the feature with respect to an infinitesimal perturbation of $\theta_j$ via 
	$\partial  h(\bmtheta) / \partial \theta_j$.
	Sensitivity to changes in a collection of parameters can be handled in an analogous fashion.  
	
	As in most calibration problems, we do not know the true value of $\bmtheta$ for the single condition where we have data from physical experimentation.  Rather, our knowledge of the parameter under the experimental treatment  is captured in the posterior distribution $f(\bmtheta|\bmx)$ given observed data $\bmx$.  We seek the related distribution for the parameters if the biologist were to intervene with an experimental manipulation, which alters some parameters in $\bmtheta$.  We call this distribution the \textit{intervention posterior}, $f_I(\bmtheta|\bmx)$. 
	Partitioning the parameter vector $\bmtheta$ into two parts, $\bmtheta = (\bmeta, \bmxi)$, we write the posterior density as $f(\bmtheta|\bmx) = f_{\eta}(\bmeta|\bmx) \cdot f_{\xi}(\bmxi|\bmeta, \bmx)$. 
	Paralleling the situation where the system parameters are known, we consider a scale change for $\bmxi$,
	which gives 
	$f_I(\bmtheta|\bmx) = f_{\eta}(\bmeta|\bmx) \cdot f_{I,\xi}(\bmxi|\bmeta, \bmx)$. 
	For example, 
	in the event that $\bmxi = \theta_j$, a scale change on $\theta_j$ by $\alpha$ would take $\bmtheta$ to $\bmnu_{j,\alpha}$ in \eqref{eqn:intervened-sample}, leaving the part corresponding to $\bmeta$ unchanged.
	
	The intervention posterior gives us access to the distribution induced on $h(\bmnu_{j,\alpha})$ by manipulating the experimental conditions to impact $\bmtheta$. This can be seen from
	the expression of the intervention posterior mean,  
	
	\vspace{-3.5em}
	\begin{align}
		&\int
		h(\bmnu_{j,\alpha}(\bmtheta))
		f(\bmnu_{j,\alpha}(\bmtheta)| \text{\bmx} ) 
		\ \diff \bmnu_{j,\alpha}(\bmtheta) 
		= \frac{1}{\alpha} \int h(\bmnu_{j,\alpha}(\bmtheta))
		f(\bmtheta|\textrm{\bmx}) \diff \bmtheta , \label{eq:intervention-expression}
	\end{align}
	\vspace{-3.5em}
	
	\noindent which allows us to run MCMC with the original posterior as the target, map the drawn $\bmtheta$ to $\bmnu_{j,\alpha}(\bmtheta)$, and compute $h(\bmnu_{j,\alpha})$ for eventual summarization and estimation.  
	
	Given the samples $\{ \bmtheta^{(b)}: b=1,\ldots,B \}$  directly drawn from the posterior distribution $f(\bmtheta|\bmx)$ using  MCMC, define $\bmnu_{j,\alpha}^{(b)}
	= \left( \theta_{1}^{(b)}, \ldots, \theta_{j-1}^{(b)}, \alpha \theta_{j}^{(b)},  \theta_{j+1}^{(b)}, \ldots, \theta_{p}^{(b)} \right)$ with a pre-specified set of $\alpha$ near 1 under the intervention posterior.  A full suite of inferences is available through $h(\bmnu_{j,\alpha}^{(b)})$ for each $j$, with varying values of $\alpha$. One straightforward way to estimate  \eqref{eq:intervention-expression} is to use
	$(1/B)\sum_{b=1}^{B} h(\bmnu_{j,\alpha}^{(b)})$
	with which we can assess the average change of the outcome $h(\cdot)$ as we decrease or increase the $j$-th parameter by $100(1-\alpha)\%$. 
	Other inferences, such as variances, quantiles, or credible intervals, can be obtained similarly. Precautions should be taken here, including measures such as examining the failure rate of model evaluations, as intervention may change the parameters beyond the range where the model is valid.
	
	The intervention posterior is closely related to the importance link function transformation of \cite{maceachern2000importance}. Their key assumption is that estimated parameters are drawn from a sampling distribution that is different from a target distribution, and a known link function is used to map the sampled values from the sampling distribution to those of the target distribution. 
	In our intervention posterior framework, the closed form of a link function is not available, yet the ODE model operates as the 
	importance link function, mapping the posterior sample space to the intervention sample space. 
	
	Sensitivity analysis has been studied in the literature on computer experiments, typically in one of two forms: a local sensitivity approach or a global one \citep{saltelli2000sensitivity, saltelli2004global,saltelli2008global,francom2018sensitivity, francom2020bass}. The approach that is closest to ours is the Accumulated Local Effects \citep[ALE, ][]{apley2020visualizing} approach. ALE calculates a set of local effects and accumulates them to provide insight into a plausible global effect. This contrasts with the intervention posterior which incorporates uncertainty in the parameters by directly changing the posterior samples. The two methods follow from distinct conceptual frameworks.  
	
	The discussion of the intervention posterior so far assumes that 
	the posterior samples are drawn from standard Metropolis-Hastings. 
	When a GMSS is used, 
	the summary under the intervention posterior can be computed as
	$(1/B) \sum_{b=1}^{B} \sum_{m=1}^M w_m^{(b)}  h(\bmnu_{m,j,\alpha}^{(b)})$    
	where $h(\bmnu_{m,j,\alpha}^{(b)})$ is obtained by running the ODE model with 
	the $j$-th variable of $\bmtheta_m^{(b)}$ of the $m$-th multiset element ``intervened''-- scaled by $\alpha$.
	
	\vspace{-2em}
	\section{Calibration of a Three-variable Biological Oscillator} \label{sec:case-study}
	In this section, we present our proposed calibration approach applied to the physical experiments and the computer simulator for the circadian cycle of \emph{N.crassa}, whose measurements are displayed in Figure \ref{fig:physical-experiment}. 
	
	\vspace{-1em}
	\subsection{Application of the GMSS}
	
	We applied the GMSS introduced in Section \ref{subsec:gmss} with $M=20$ multiset elements. To find a set of working instrumental densities for the GMSS, 
	we conducted prognostic experiments as described in Section \ref{subsec:prognostic-experiments} with 100 independent batches of orthogonal array-based Latin hypercube designs using a full $3^9$ factorial design, which took 17 hours using 36 Xeon E5-2680 cores. The parameters used in the search process were $d_0 = 4$, $n_{\min} = 0$, and $l_{\min} = -150$. That is, any experiments  with log-likelihood higher than $-150$ were deemed successful, and any level combination in the $3^9$ design that matches in four columns or more with a single successful run were given $\rho_1 = 1.0$, and $\rho_0 = 0.1$ otherwise, where 13\% ($=$2,585/19,683) of  the $3^9$ factorial designs  level combinations were designated as the primary search areas.
	
\begin{figure}
		\centering
		\includegraphics[width=0.4\textwidth]{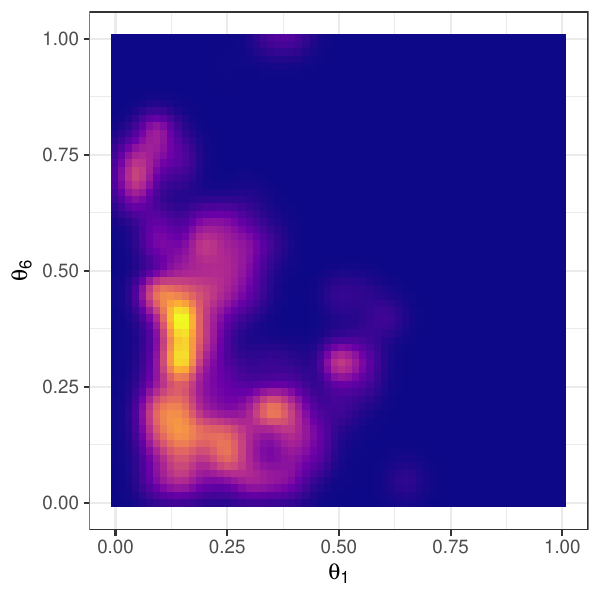} \includegraphics[width=0.4\textwidth]{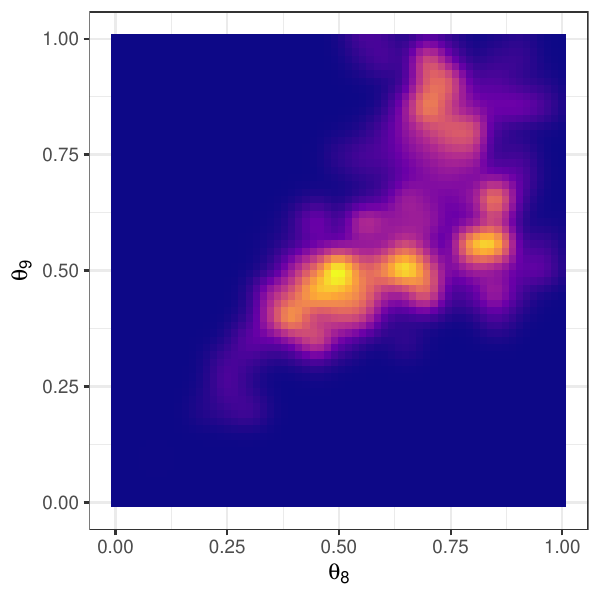}
		\caption{Bivariate posterior distributions of $(\theta_1, \theta_6)$ and $(\theta_8, \theta_9)$ obtained from GMSS.}
		\label{fig:bivariate}
	\end{figure}

	With the GMSS, the chain successfully converged and mixed well in $B=$200,000 iterations. In contrast, an ordinary Metropolis-Hastings chain failed to reach a stationary distribution even after $B=4$ million iterations. 
	Figure \ref{fig:bivariate} presents a pair of heatmaps of  bivariate posterior distribution obtained from the GMSS. Standard Metropolis-Hastings is hindered by the complexity of the distribution, while GMSS successfully explores to the target distribution with multiple modes as shown in the figures above. 
	
	\vspace{-1em}
	\subsection{Sensitivity Analysis with Intervention Posterior} \label{subsec:SA-intervention}
	We conducted a sensitivity analysis using the intervention posterior introduced in Section \ref{sec:intervention-posterior} with GMSS samples. 
	\begin{figure}
		\centering
		\includegraphics[width=.6\textwidth]{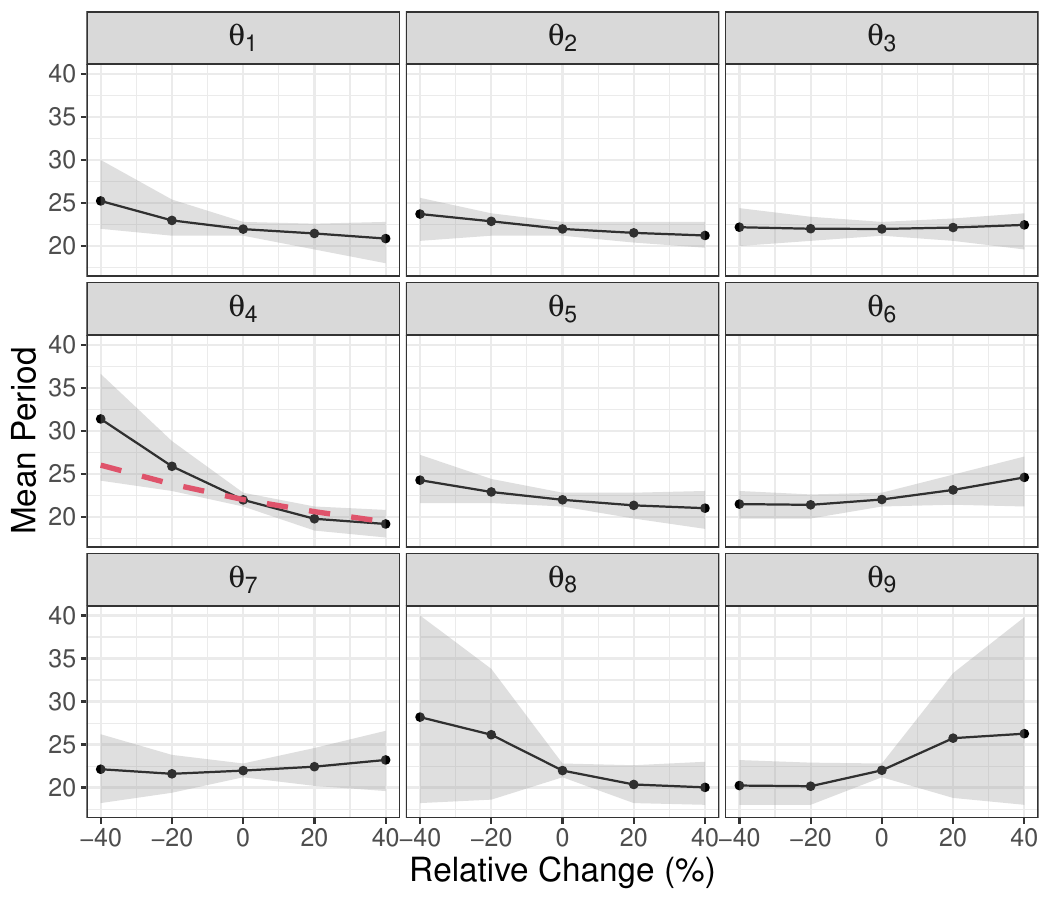}
		\caption{Sensitivity analysis of all nine variables using the intervention posterior approach, where the solid dots and line indicate the posterior mean of the circadian cycle's period, and the shaded areas represent an 80\% credible interval. The sensitivity analysis for $\theta_4$ from \cite{caicedo2015robustness} is overlaid with the dashed line for the corresponding panel.}
		\label{fig:global-sensitivity}
	\end{figure}
	
	Figure  \ref{fig:global-sensitivity} 
	shows the posterior sensitivity for period of the circadian cycle, evaluated with 
	$\alpha = (0.6, 0.8, 1.0, 1.2, 1.4)$.   The solid dots indicate the mean circadian cycle while the shaded areas represent 80\% credible intervals.  The strong slope in the plot for $\theta_4$ shows that this parameter affects periodicity the most. The parameters $\theta_2$, $\theta_3$, and $\theta_6$  seem inert, whereas $\theta_1$, $\theta_5$, and $\theta_7$ show a modest effect on period. The parameters $\theta_8$ and $\theta_9$ show a comparable effect, yet with credible intervals of great width, which indicates that the effect on period varies substantially, depending on the values of the other parameters. Another note regarding $\theta_8$ and $\theta_9$ is that the computer model often fails to converge when the parameter values are changed from its posterior values, as reported in Table S1 of the Supplement.
	
\begin{figure}
		\centering
		\includegraphics[width=0.76\textwidth]{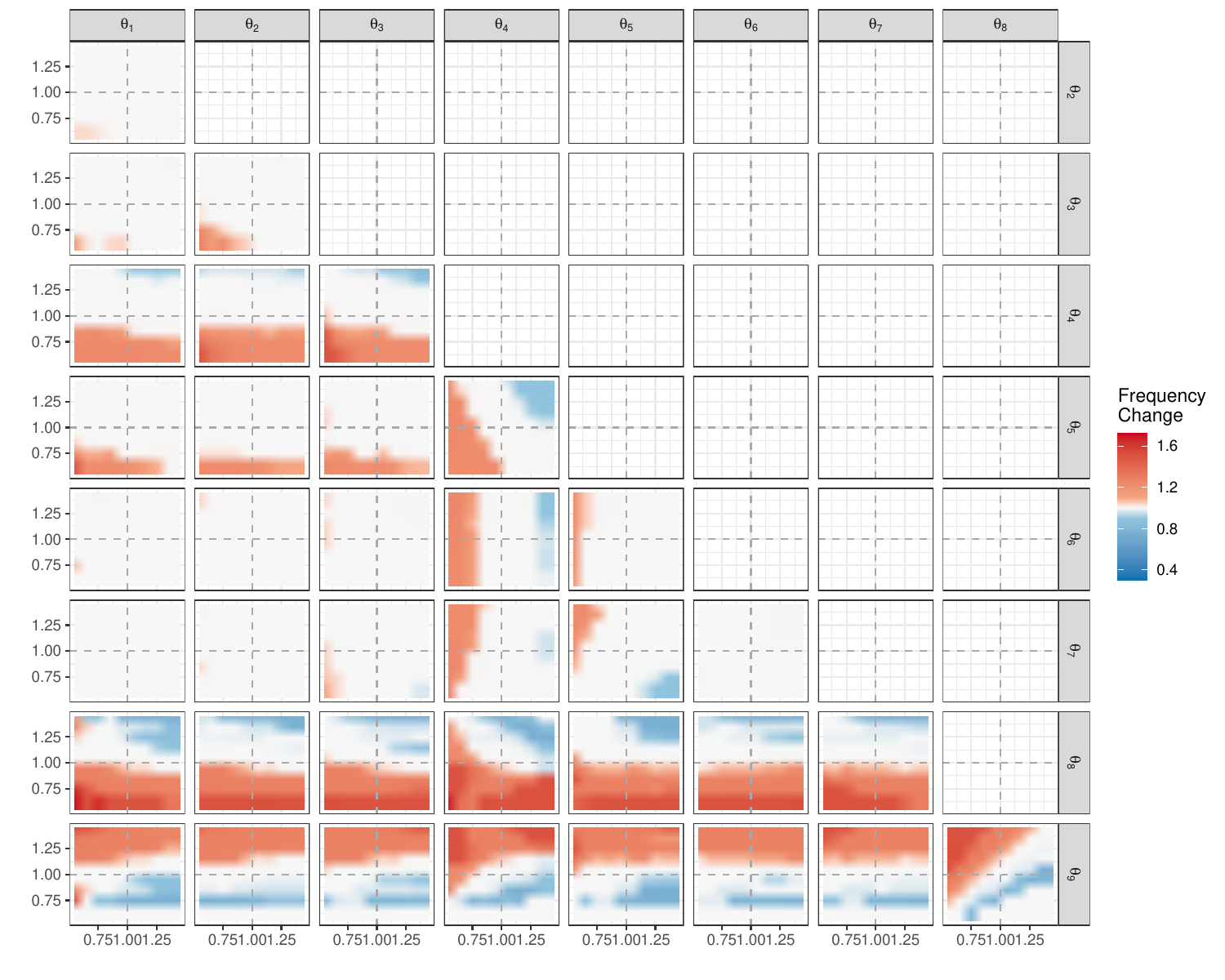}
		\caption{Sensitivity analysis of all pairs of variables using the intervention posterior approach, where color represents the relative change in the period of the circadian cycle.}
		\label{fig:bivariate-global-sensitivity}
	\end{figure}

	To illustrate the scientific contribution of our approach, we superimposed the result from \cite{caicedo2015robustness} over our results in Figure \ref{fig:global-sensitivity}.
	\cite{caicedo2015robustness} manually  explored the parameter space  and concluded that $\theta_4$ had the most significant impact  on the period. Then the impact of intervention was explored by changing its value while holding other parameters fixed and then examining the change in the period of the circadian cycle as a function of the varying $\theta_4$ value. 
	Both approaches show a common overall trend, however, \cite{caicedo2015robustness} underestimates the sensitivity as their single estimate is much flatter than the curve ofintervention posterior means.  
	In contrast, our approach relies on the intervention posterior to examine the sensitivity of the circadian cycle's period to changes in $\theta_4$.  Since the posterior distribution shows considerable dispersion and our inferential target is a highly nonlinear function of the parameters, 
	selecting a single parameter value about which to consider perturbations misses the fact that the impact of perturbation could be very different for other parameter values.  
	Figure \ref{fig:bivariate-global-sensitivity} shows a heat map for posterior mean period, considering perturbation of two parameters at a time.  Each parameter is perturbed with $\alpha$ ranging from $0.6$ to $1.4$.  The size of the perturbation is given on the axes of the figure.  The plotted (heat) value is the relative change in period.  These plots are effective in conveying the sensitivity of period to perturbation and in conveying interactions or the lack thereof.  
	For instance, lowering the value of $\theta_4$ tends to increases the period, while it can be compensated for by raising the value of $\theta_5$.
	
	\vspace{-1.5em}
	
	\section{Discussion} \label{sec:discussion}
	In this work, we address a challenging scientific problem in analyzing oscillatory phenomena in biological systems. The proposed computational framework provides a rigorous methodology for model calibration, uncertainty quantification, and sensitivity analysis for biological studies investigating periodic phenotypes.
	The GMSS, equipped with instrumental densities obtained from prognostic experiments, successfully finds posterior densities that lie in a thin high-dimensional manifold. The intervention posterior approach is devised for an intuitive  yet rigorous sensitivity analysis to better utilize the result from Bayesian calibration. 
	Our case study presents an overhaul of the analysis of the model from \cite{caicedo2015robustness}.
	Placing the analysis in a statistical framework generates more insights, 
	which is expected to attract wide research interest from 
	biology communities studying oscillatory phenomena ranging from cell cycle to circadian rhythms. 
	
	Like other standard MCMC methods, the dimensionality that GMSS can handle is limited by the computation time and resources available for the analyst. If one uses GMSS with a uniform instrumental density, the number of multiset elements required for finding unknown multiple modes increases exponentially as the dimension of parameter space increases. In higher dimensions, it is more crucial to find appropriate instrumental densities using prognostic experiments as discussed in Section \ref{subsec:prognostic-experiments}.
	
	The overall shape of a curve produced by the harmonic representation is determined by the amplitude \emph{and} the phase of each individual basis component. In our modeling approach, we match only the amplitude $a_k^2+b_k^2$ of each individual basis component. This can lead to the acceptance of parameter values that produce ODE model output differing from the field data not only in the phase of the entire curve but also in its specific shape, which is determined by the phase alignment of each basis component. As a result, the accepted parameters may generate ODE curves with slightly different shapes, arising particularly from the higher-frequency basis components. Despite this limitation, our approach works well for the current problem because most of the output can be effectively approximated by the primary periodic component (22-hour cycle, $k=3$) and the overall damping components ($k=1, 2$), which align with the field data. However, our approach would require significant modification to handle more complex periodic behaviors. In such cases, it may be necessary to introduce an additional layer to the model to account for phase alignment among basis components. The development of such an approach, however, would require considerable effort, and we leave it as a potential direction for future research.
	
	We conclude with remarks on potential future research topics. First, the intervention posterior is a useful tool for sensitivity analysis, providing access to counterfactual scenarios through biological computer models. It is closely related to the concept of expected effect size; hence sample size calculation can be an immediate research topic to follow. 
	Second, we consider a fully deterministic model based on differential equations.  
	When studying a long series of data, a model with a stochastic evolution of the system can be more useful because the governing system itself can evolve over time. Calibrating such stochastic models requires the development of a new calibration approach. 
	The intervention posterior may need a suitable modification for those cases.

	\section*{Supplementary Materials}
	The supplementary materials contain the following:
	(i) MCMC steps for the standard Metropolis-within-Gibbs algorithm for the proposed hierarchical Bayesian model in \eqref{eqn:BayesianModel}, (ii) MCMC steps for the Generalized Multiset Sampler for the proposed hierarchical Bayesian model in \eqref{eqn:BayesianModel}, (iii) a figure showing an example of a computer model output from a posterior sample, compared with the field experiment results, (iv) a figure displaying five accepted simulation runs to highlight the phase shift and minor amplitude variations, (v) a figure showing the trace of the leading multiset element. Additionally, we have included the code to reproduce Figure \ref{fig:global-sensitivity}, along with the data files for field data experiments. For detailed instructions on running the code and associated data files, please refer to the \texttt{readme} document.
	
	\section*{Acknowledgments}
	The authors are deeply grateful to the editor, the associate editor, and three referees for their reviews, which significantly improved the article.

	\section*{Additional information}
	\subsection*{Funding}
	These authors gratefully acknowledge funding from PSC-CUNY Grant, NSF DMS-2413823, New Faculty Startup Fund from Seoul National University 326-20240027, NIH R01DK117005, University of Cincinnati College of Medicine Bridge Funding,  University of Cincinnati Cancer Center Pilot Funding, and Nano \& Material Technology Development Program through the National Research Foundation of Korea funded by Ministry of Science and ICT (RS-2024-00451383).

	\section*{Disclosure Statement}
	No potential conflict of interest was reported by the authors.
	
	\bibliographystyle{jasa3}
	\bibliography{reference}

\begin{thebibliography}{49}
\newcommand{\enquote}[1]{``#1''}
\expandafter\ifx\csname natexlab\endcsname\relax\def\natexlab#1{#1}\fi
\expandafter\ifx\csname url\endcsname\relax
  \def\url#1{{\tt #1}}\fi
\expandafter\ifx\csname urlprefix\endcsname\relax\def\urlprefix{URL }\fi

\bibitem[\protect\citeauthoryear{Apley and Zhu}{Apley and
  Zhu}{2020}]{apley2020visualizing}
Apley, D.~W. and Zhu, J. (2020), \enquote{Visualizing the effects of predictor
  variables in black box supervised learning models,} {\em Journal of the Royal
  Statistical Society Series B: Statistical Methodology\/}, 82, 1059--1086.

\bibitem[\protect\citeauthoryear{Bayarri, Berger, Cafeo, Garcia-Donato, Liu,
  Palomo, Parthasarathy, Paulo, Sacks, and Walsh}{Bayarri
  et~al.}{2007}]{bayarri2007computer}
Bayarri, M., Berger, J., Cafeo, J., Garcia-Donato, G., Liu, F., Palomo, J.,
  Parthasarathy, R., Paulo, R., Sacks, J., and Walsh, D. (2007),
  \enquote{Computer model validation with functional output,} {\em Annals of
  Statistics\/}, 35, 1874--1906.

\bibitem[\protect\citeauthoryear{Bell-Pedersen, Cassone, Earnest, Golden,
  Hardin, Thomas, and Zoran}{Bell-Pedersen et~al.}{2005}]{bell2005circadian}
Bell-Pedersen, D., Cassone, V.~M., Earnest, D.~J., Golden, S.~S., Hardin,
  P.~E., Thomas, T.~L., and Zoran, M.~J. (2005), \enquote{Circadian rhythms
  from multiple oscillators: lessons from diverse organisms,} {\em Nature
  Reviews Genetics\/}, 6, 544--556.

\bibitem[\protect\citeauthoryear{Bellman, Kim, Lim, and Hong}{Bellman
  et~al.}{2018}]{bellman2018modeling}
Bellman, J., Kim, J.~K., Lim, S., and Hong, C.~I. (2018), \enquote{Modeling
  reveals a key mechanism for light-dependent phase shifts of {\it Neurospora}
  circadian rhythms,} {\em Biophysical Journal\/}, 115, 1093--1102.

\bibitem[\protect\citeauthoryear{Bhat, Haran, Olson, and Keller}{Bhat
  et~al.}{2012}]{bhat2013inferring}
Bhat, K., Haran, M., Olson, R., and Keller, K. (2012), \enquote{{Inferring
  likelihoods and climate system characteristics from climate models and
  multiple tracers},} {\em Environmetrics\/}, 23, 345--362.

\bibitem[\protect\citeauthoryear{Caicedo-Casso, Kang, Lim, and
  Hong}{Caicedo-Casso et~al.}{2015}]{caicedo2015robustness}
Caicedo-Casso, A., Kang, H.-W., Lim, S., and Hong, C.~I. (2015),
  \enquote{Robustness and period sensitivity analysis of minimal models for
  biochemical oscillators,} {\em Scientific Reports\/}, 5, 1--15.

\bibitem[\protect\citeauthoryear{Chang, Haran, Applegate, and Pollard}{Chang
  et~al.}{2016}]{chang2015binary}
Chang, W., Haran, M., Applegate, P., and Pollard, D. (2016),
  \enquote{{Calibrating an ice sheet model using high-dimensional binary
  spatial data},} {\em Journal of American Statistical Association\/}, 111,
  57--72.

\bibitem[\protect\citeauthoryear{Chang, Haran, Olson, and Keller}{Chang
  et~al.}{2014}]{chang2014fast}
Chang, W., Haran, M., Olson, R., and Keller, K. (2014), \enquote{Fast
  dimension-reduced climate model calibration and the effect of data
  aggregation,} {\em Annals of Applied Statistics\/}, 8, 649--673.

\bibitem[\protect\citeauthoryear{Dunlap and Loros}{Dunlap and
  Loros}{2017}]{dunlap2017making}
Dunlap, J.~C. and Loros, J.~J. (2017), \enquote{Making time: conservation of
  biological clocks from fungi to animals,} {\em Microbiology Spectrum\/}, 5.

\bibitem[\protect\citeauthoryear{Francom and Sans{\'o}}{Francom and
  Sans{\'o}}{2020}]{francom2020bass}
Francom, D. and Sans{\'o}, B. (2020), \enquote{BASS: An R package for fitting
  and performing sensitivity analysis of Bayesian adaptive spline surfaces,}
  {\em Journal of Statistical Software\/}, 94.

\bibitem[\protect\citeauthoryear{Francom, Sans{\'o}, Bulaevskaya, Lucas, and
  Simpson}{Francom et~al.}{2019}]{francom2019inferring}
Francom, D., Sans{\'o}, B., Bulaevskaya, V., Lucas, D., and Simpson, M. (2019),
  \enquote{Inferring atmospheric release characteristics in a large computer
  experiment using Bayesian adaptive splines,} {\em Journal of the American
  Statistical Association\/}, 114, 1450--1465.

\bibitem[\protect\citeauthoryear{Francom, Sans{\'o}, Kupresanin, and
  Johannesson}{Francom et~al.}{2018}]{francom2018sensitivity}
Francom, D., Sans{\'o}, B., Kupresanin, A., and Johannesson, G. (2018),
  \enquote{Sensitivity analysis and emulation for functional data using
  Bayesian adaptive splines,} {\em Statistica Sinica\/}, 791--816.

\bibitem[\protect\citeauthoryear{Francom, Tucker, Huerta, Shuler, and
  Ries}{Francom et~al.}{2023}]{francom2023elastic}
Francom, D., Tucker, J.~D., Huerta, G., Shuler, K., and Ries, D. (2023),
  \enquote{Elastic Bayesian model calibration,} {\em arXiv preprint
  arXiv:2305.08834\/}, 1--36.

\bibitem[\protect\citeauthoryear{Gallego, Eide, Woolf, Virshup, and
  Forger}{Gallego et~al.}{2006}]{gallego2006opposite}
Gallego, M., Eide, E.~J., Woolf, M.~F., Virshup, D.~M., and Forger, D.~B.
  (2006), \enquote{An opposite role for {\it tau} in circadian rhythms revealed
  by mathematical modeling,} {\em Proceedings of the National Academy of
  Sciences\/}, 103, 10618--10623.

\bibitem[\protect\citeauthoryear{Geyer}{Geyer}{1991}]{geyer1991markov}
Geyer, C.~J. (1991), \enquote{Markov chain Monte Carlo maximum likelihood,} in
  Keramides, E.~M. (editor), {\em Computing Science and Statistics: Proceedings
  of the 23rd Symposium on the Interface\/}, Fairfax Station, Va.: Interface
  Foundation.

\bibitem[\protect\citeauthoryear{Gooch, Johnson, Bourne, Nix, Maas, Fox, Loros,
  Larrondo, and Dunlap}{Gooch et~al.}{2014}]{gooch2014kinetic}
Gooch, V.~D., Johnson, A.~E., Bourne, B.~J., Nix, B.~T., Maas, J.~A., Fox,
  J.~A., Loros, J.~J., Larrondo, L.~F., and Dunlap, J.~C. (2014), \enquote{A
  kinetic study of the effects of light on circadian rhythmicity of the {\it
  frq} promoter of {\it Neurospora crassa},} {\em Journal of Biological
  Rhythms\/}, 29, 38--48.

\bibitem[\protect\citeauthoryear{Gramacy, Bingham, Holloway, Grosskopf, Kuranz,
  Rutter, Trantham, and Drake}{Gramacy et~al.}{2015}]{gramacy2015calibrating}
Gramacy, R.~B., Bingham, D., Holloway, J.~P., Grosskopf, M.~J., Kuranz, C.~C.,
  Rutter, E., Trantham, M., and Drake, R.~P. (2015), \enquote{Calibrating a
  large computer experiment simulating radiative shock hydrodynamics,} {\em
  Annals of Applied Statistics\/}, 9, 1141--1168.

\bibitem[\protect\citeauthoryear{Gu}{Gu}{2019}]{gu2019jointly}
Gu, M. (2019), \enquote{Jointly robust prior for Gaussian stochastic process in
  emulation, calibration and variable selection,} {\em Bayesian Analysis\/},
  14, 857--885.

\bibitem[\protect\citeauthoryear{Hartley}{Hartley}{1949}]{hartley1949tests}
Hartley, H.~O. (1949), \enquote{Tests of significance in harmonic analysis,}
  {\em Biometrika\/}, 36, 194--201.

\bibitem[\protect\citeauthoryear{Higdon, Gattiker, Williams, and
  Rightley}{Higdon et~al.}{2008}]{higdon2008computer}
Higdon, D., Gattiker, J., Williams, B., and Rightley, M. (2008),
  \enquote{Computer model calibration using high-dimensional output,} {\em
  Journal of the American Statistical Association\/}, 103, 570--583.

\bibitem[\protect\citeauthoryear{Hwang, Barut, and Yeo}{Hwang
  et~al.}{2018}]{hwang2018statistical}
Hwang, Y., Barut, E., and Yeo, K. (2018), \enquote{Statistical-physical
  estimation of pollution emission,} {\em Statistica Sinica\/}, 28, 921--940.

\bibitem[\protect\citeauthoryear{Hwang, He, and Qian}{Hwang
  et~al.}{2016}]{hwang2016sliced}
Hwang, Y., He, X., and Qian, P.~Z. (2016), \enquote{Sliced orthogonal
  array-based Latin hypercube designs,} {\em Technometrics\/}, 58, 50--61.

\bibitem[\protect\citeauthoryear{Karagiannis and Lin}{Karagiannis and
  Lin}{2017}]{karagiannis2017bayesian}
Karagiannis, G. and Lin, G. (2017), \enquote{On the Bayesian calibration of
  computer model mixtures through experimental data, and the design of
  predictive models,} {\em Journal of Computational Physics\/}, 342, 139--160.

\bibitem[\protect\citeauthoryear{Kennedy and O'Hagan}{Kennedy and
  O'Hagan}{2001}]{kennedy2001bayesian}
Kennedy, M.~C. and O'Hagan, A. (2001), \enquote{Bayesian calibration of
  computer models,} {\em Journal of the Royal Statistical Society: Series B
  (Statistical Methodology)\/}, 63, 425--464.

\bibitem[\protect\citeauthoryear{Kim and MacEachern}{Kim and
  MacEachern}{2015}]{kim2015generalized}
Kim, H.~J. and MacEachern, S.~N. (2015), \enquote{The generalized multiset
  sampler,} {\em Journal of Computational and Graphical Statistics\/}, 24,
  1134--1154.

\bibitem[\protect\citeauthoryear{Kirkpatrick}{Kirkpatrick}{1984}]{kirkpatrick1984optimization}
Kirkpatrick, S. (1984), \enquote{Optimization by simulated annealing:
  Quantitative studies,} {\em Journal of Statistical Physics\/}, 34, 975--986.

\bibitem[\protect\citeauthoryear{Landa, Mossman, Whitaker, Rapti, and
  Clifton}{Landa et~al.}{2022}]{landa2022phage}
Landa, K.~J., Mossman, L.~M., Whitaker, R.~J., Rapti, Z., and Clifton, S.~M.
  (2022), \enquote{Phage--antibiotic synergy inhibited by temperate and chronic
  virus competition,} {\em Bulletin of Mathematical Biology\/}, 84, 54.

\bibitem[\protect\citeauthoryear{Lee, Haran, Fuller, Pollard, and Keller}{Lee
  et~al.}{2020}]{lee2020fast}
Lee, B.~S., Haran, M., Fuller, R.~W., Pollard, D., and Keller, K. (2020),
  \enquote{A fast particle-based approach for calibrating a 3-D model of the
  Antarctic ice sheet,} {\em Annals of Applied Statistics\/}, 14, 605--634.

\bibitem[\protect\citeauthoryear{Leman, Chen, and Lavine}{Leman
  et~al.}{2009}]{leman2009multiset}
Leman, S.~C., Chen, Y., and Lavine, M. (2009), \enquote{The multiset sampler,}
  {\em Journal of the American Statistical Association\/}, 104, 1029--1041.

\bibitem[\protect\citeauthoryear{Liu, Bayarri, and Berger}{Liu
  et~al.}{2009}]{liu2009modularization}
Liu, F., Bayarri, M., and Berger, J. (2009), \enquote{Modularization in
  Bayesian analysis, with emphasis on analysis of computer models,} {\em
  Bayesian Analysis\/}, 4, 119--150.

\bibitem[\protect\citeauthoryear{Liu, Liang, and Wong}{Liu
  et~al.}{2000}]{liu2000multiple}
Liu, J.~S., Liang, F., and Wong, W.~H. (2000), \enquote{The multiple-try method
  and local optimization in Metropolis sampling,} {\em Journal of the American
  Statistical Association\/}, 95, 121--134.

\bibitem[\protect\citeauthoryear{Liu, Chen, Caicedo-Casso, Cui, Du, He, Lim,
  Kim, Hong, and Liu}{Liu et~al.}{2019}]{liu2019frq}
Liu, X., Chen, A., Caicedo-Casso, A., Cui, G., Du, M., He, Q., Lim, S., Kim,
  H.~J., Hong, C.~I., and Liu, Y. (2019), \enquote{FRQ-CK1 interaction
  determines the period of circadian rhythms in {\it Neurospora},} {\em Nature
  Communications\/}, 10, 1--13.

\bibitem[\protect\citeauthoryear{MacEachern and Peruggia}{MacEachern and
  Peruggia}{2000}]{maceachern2000importance}
MacEachern, S.~N. and Peruggia, M. (2000), \enquote{Importance link function
  estimation for Markov chain Monte Carlo methods,} {\em Journal of
  Computational and Graphical Statistics\/}, 9, 99--121.

\bibitem[\protect\citeauthoryear{Marinari and Parisi}{Marinari and
  Parisi}{1992}]{marinari1992simulated}
Marinari, E. and Parisi, G. (1992), \enquote{Simulated tempering: a new Monte
  Carlo scheme,} {\em Europhysics Letters\/}, 19, 451--458.

\bibitem[\protect\citeauthoryear{Orme{\~n}o and General}{Orme{\~n}o and
  General}{2024}]{ormeno2024convergence}
Orme{\~n}o, F. and General, I.~J. (2024), \enquote{Convergence and equilibrium
  in molecular dynamics simulations,} {\em Communications Chemistry\/}, 7, 26.

\bibitem[\protect\citeauthoryear{Peifer and Timmer}{Peifer and
  Timmer}{2007}]{peifer2007parameter}
Peifer, M. and Timmer, J. (2007), \enquote{Parameter estimation in ordinary
  differential equations for biochemical processes using the method of multiple
  shooting,} {\em IET Systems Biology\/}, 1, 78--88.

\bibitem[\protect\citeauthoryear{Saltelli}{Saltelli}{2004}]{saltelli2004global}
Saltelli, A. (2004), \enquote{Global sensitivity analysis: an introduction,} in
  {\em Proc. 4th International Conference on Sensitivity Analysis of Model
  Output (SAMO’04)\/}, volume~27, Citeseer.

\bibitem[\protect\citeauthoryear{Saltelli, Campolongo, and Tarantola}{Saltelli
  et~al.}{2000}]{saltelli2000sensitivity}
Saltelli, A., Campolongo, F., and Tarantola, S. (2000), \enquote{{Sensitivity
  anaysis as an ingredient of modeling},} {\em Statistical Science\/}, 15,
  377--395.

\bibitem[\protect\citeauthoryear{Saltelli, Ratto, Andres, Campolongo, Cariboni,
  Gatelli, Saisana, and Tarantola}{Saltelli et~al.}{2008}]{saltelli2008global}
Saltelli, A., Ratto, M., Andres, T., Campolongo, F., Cariboni, J., Gatelli, D.,
  Saisana, M., and Tarantola, S. (2008), {\em {Global Sensitivity Analysis: the
  Primer}\/}, John Wiley \& Sons.

\bibitem[\protect\citeauthoryear{Salter, Williamson, Scinocca, and
  Kharin}{Salter et~al.}{2019}]{salter2019uncertainty}
Salter, J.~M., Williamson, D.~B., Scinocca, J., and Kharin, V. (2019),
  \enquote{Uncertainty quantification for computer models with spatial output
  using calibration-optimal bases,} {\em Journal of the American Statistical
  Association\/}, 114, 1800--1814.

\bibitem[\protect\citeauthoryear{Santner, Williams, Notz, and Williams}{Santner
  et~al.}{2018}]{santner2003design}
Santner, T.~J., Williams, B.~J., Notz, W.~I., and Williams, B.~J. (2018), {\em
  {The Design and Analysis of Computer Experiments}\/}, New York: Springer, 2nd
  edition.

\bibitem[\protect\citeauthoryear{Sassone-Corsi, Young, and Reddy}{Sassone-Corsi
  et~al.}{2018}]{Corsietal2018}
Sassone-Corsi, P., Young, M.~W., and Reddy, A.~B. (2018), {\em Circadian
  Rhythms: A Subject Collection From Cold Spring Harbor Perspectives in
  Biology\/}, New York: Cold Spring Harbor Laboratory Press.

\bibitem[\protect\citeauthoryear{Sung, Hung, Rittase, Zhu, and Wu}{Sung
  et~al.}{2020}]{sung2020calibration}
Sung, C.-L., Hung, Y., Rittase, W., Zhu, C., and Wu, C. F.~J. (2020),
  \enquote{Calibration for computer experiments with binary responses and
  application to cell adhesion study,} {\em Journal of the American Statistical
  Association\/}, 115, 1664--1674.

\bibitem[\protect\citeauthoryear{Swendsen and Wang}{Swendsen and
  Wang}{1986}]{swendsen1986replica}
Swendsen, R.~H. and Wang, J.-S. (1986), \enquote{Replica Monte Carlo simulation
  of spin-glasses,} {\em Physical Review Letters\/}, 57, 2607--2609.

\bibitem[\protect\citeauthoryear{Tang}{Tang}{1993}]{tang1993orthogonal}
Tang, B. (1993), \enquote{Orthogonal array-based Latin hypercubes,} {\em
  Journal of the American Statistical Association\/}, 88, 1392--1397.

\bibitem[\protect\citeauthoryear{Tsai, Choi, Ma, Pomerening, Tang, and
  Ferrell}{Tsai et~al.}{2008}]{tsai2008robust}
Tsai, T. Y.-C., Choi, Y.~S., Ma, W., Pomerening, J.~R., Tang, C., and Ferrell,
  J.~E. (2008), \enquote{Robust, tunable biological oscillations from
  interlinked positive and negative feedback loops,} {\em Science\/}, 321,
  126--129.

\bibitem[\protect\citeauthoryear{Tu, Kudlicki, Rowicka, and McKnight}{Tu
  et~al.}{2005}]{tu2005logic}
Tu, B.~P., Kudlicki, A., Rowicka, M., and McKnight, S.~L. (2005),
  \enquote{Logic of the yeast metabolic cycle: temporal compartmentalization of
  cellular processes,} {\em Science\/}, 310, 1152--1158.

\bibitem[\protect\citeauthoryear{Tuo and Wu}{Tuo and
  Wu}{2015}]{tuo2015efficient}
Tuo, R. and Wu, C.~J. (2015), \enquote{Efficient calibration for imperfect
  computer models,} {\em Annals of Statistics\/}, 43, 2331--2352.

\bibitem[\protect\citeauthoryear{Wu and Hamada}{Wu and
  Hamada}{2021}]{wu2011experiments}
Wu, C.~J. and Hamada, M.~S. (2021), {\em Experiments: Planning, Analysis, and
  Optimization\/}, New Jersey: John Wiley \& Sons, 3rd edition.

\end{thebibliography}

\end{document}